# Anomalous charge density wave in a two-dimensional superatomic superconductor

Boqin Song[1], Shuaishuai Sun[1], Zhongxu Wei[1], Xinbo Wang[1], Xiaoping Ma[1], Kaifa Luo[2], Lei Wang[1], Jun Deng[3], Xu Chen[1], Tian Qian[1], Shuya Xing[4], Zhihai Cheng[5], Jian-gang Guo[1,*], Tianping Ying[1,*], Xiaolong Chen[1,*]

1. Beijing National Laboratory for Condensed Matter Physics, Institute of Physics, Chinese Academy of Sciences, Beijing 100190, China
2. Department of Physics, University of Texas at Austin, Austin, Texas 78712, USA
3. Center for High Pressure Science and Technology Advanced Research, Beijing 100093, China
4. Research Center for Quantum Physics and Technologies, Inner Mongolia University, Hohhot 010021, China
5. Beijing Key Laboratory of Optoelectronic Functional Materials and Micro-nano Devices, Department of Physics, Renmin University of China, Beijing 100872, China

*ying@iphy.ac.cn
*jgguo@iphy.ac.cn
*xlchen@iphy.ac.cn

## Abstract

**The spatial modulation of electron density into a wave-like pattern, known as charge density wave (CDW), represents a fundamental quantum state that often coexists with superconductivity, quantum Hall states, axion insulating phases and etc. Conventional CDWs are mediated by longitudinal acoustic phonons, exhibit picometer-scale lattice distortions ($10^{-12}$-$10^{-11}$ m), and typically vanish approaching the atomic limit. Here, we report a series of anomalous CDW behaviors in the 2D superatomic superconductor $Au_6Te_{12}Se_8$. Remarkably, its CDW is governed by transverse phonons, accompanied by an extraordinarily high real-space displacement of ~ 4 Ångström. Furthermore, we observe an exotic dimensional response persisting up to micrometer-scale thickness, a regime where other materials are already considered as bulk. Through liquid helium-temperature transmission electron microscopy, ultrafast pump-probe spectroscopy and transport measurements, we demonstrate a dramatic enhancement of the CDW transition temperature ($T_{CDW}$) from <2 K in the bulk to 110 K in approaching the "superatomic limit". Our findings not only reveal novel facets of both CDW and superatomic materials, but the competition between this anomalous CDW and superconductivity opens avenues for exploring unconventional electron-phonon interactions.**

## Introduction

The essence of “more is different” manifests in the emergent properties arising from quantitative changes in material building blocks[1,2]. When the fundamental units evolve from single atoms to tailored superatoms, and chemical bonds transform into inter-cluster bonds[3-7], these additional degrees of freedom unlock extraordinary phenomena beyond the reach of conventional atomic crystals[8-12], including extraordinary-high charge carrier mean free paths and exceptionally long-lived exciton-polaron quasiparticles[13]. This paradigm shift naturally invites exploration of diverse quantum ground states in this expanded material category. Among these, CDW occupies a unique position in condensed matter physics[14-23], often serving as precursor to superconductivity and magnetic order. While both superconductivity and magnetism have already been established in superatomic materials[24-27], study on the CDW remains to be scarce, with only a tentative CDW-like feature reported in $Au_6Te_{12}Se_8$ to date[28,29].

While most superatomic compounds are nonmetallic due to their large inter-cluster distances[6,7,11,30], $Au_6Te_{12}Se_8$ represents a rare exception, exhibiting metallic conductivity at high temperatures before transition into a superconductor below 2.6 K[25]. As shown in Fig. 1a, the intralayer spacing between $Au_6Te_{12}Se_8$ superatoms (3.79 Å and 3.64 Å) are comparable to the interlayer gap of 4.5 Å, creating a unique van der Waals system with ultralow electrical anisotropy ($\rho_c/\rho_{ab} \sim 1$)[28], a stark contrast to conventional 2D materials ($\rho_c/\rho_{ab} > 100$)[31-33]. To date, CDW behavior in $Au_6Te_{12}Se_8$ has only been observed through surface electron reconstructions, with no evidence of bulk lattice distortions[29]. The conflicting transport measurements further obscure the origin of this phenomenon[25,28], casting doubts on whether superatomic CDWs emulate or transcend the landscape of their atomic counterparts.

Here, through a comprehensive investigation combining liquid-helium-temperature transmission electron microscopy (He-TEM), scanning tunneling microscopy (STM), ultrafast pump-probe spectroscopy, and transport measurements, we discover a series of anomalous CDW behaviors in $Au_6Te_{12}Se_8$. Unlike conventional CDWs that weaken approaching the atomic limit, the superatomic CDW exhibits a striking dimensional enhancement. Moreover, the observed CDW couples not to the textbook longitudinal acoustic phonons, but to transverse phonon modes, generating real-space lattice distortions 1-2 orders of magnitude larger than those in atomic systems. Detailed analysis traces these observed behaviors to the unique inter-cluster bonds. A systematic dimensional evolution approaching the “superatomic limit” and the direct competition between superatomic CDW and superconductivity is revealed for the first time.

## Results

### Evidence of lattice reconstruction

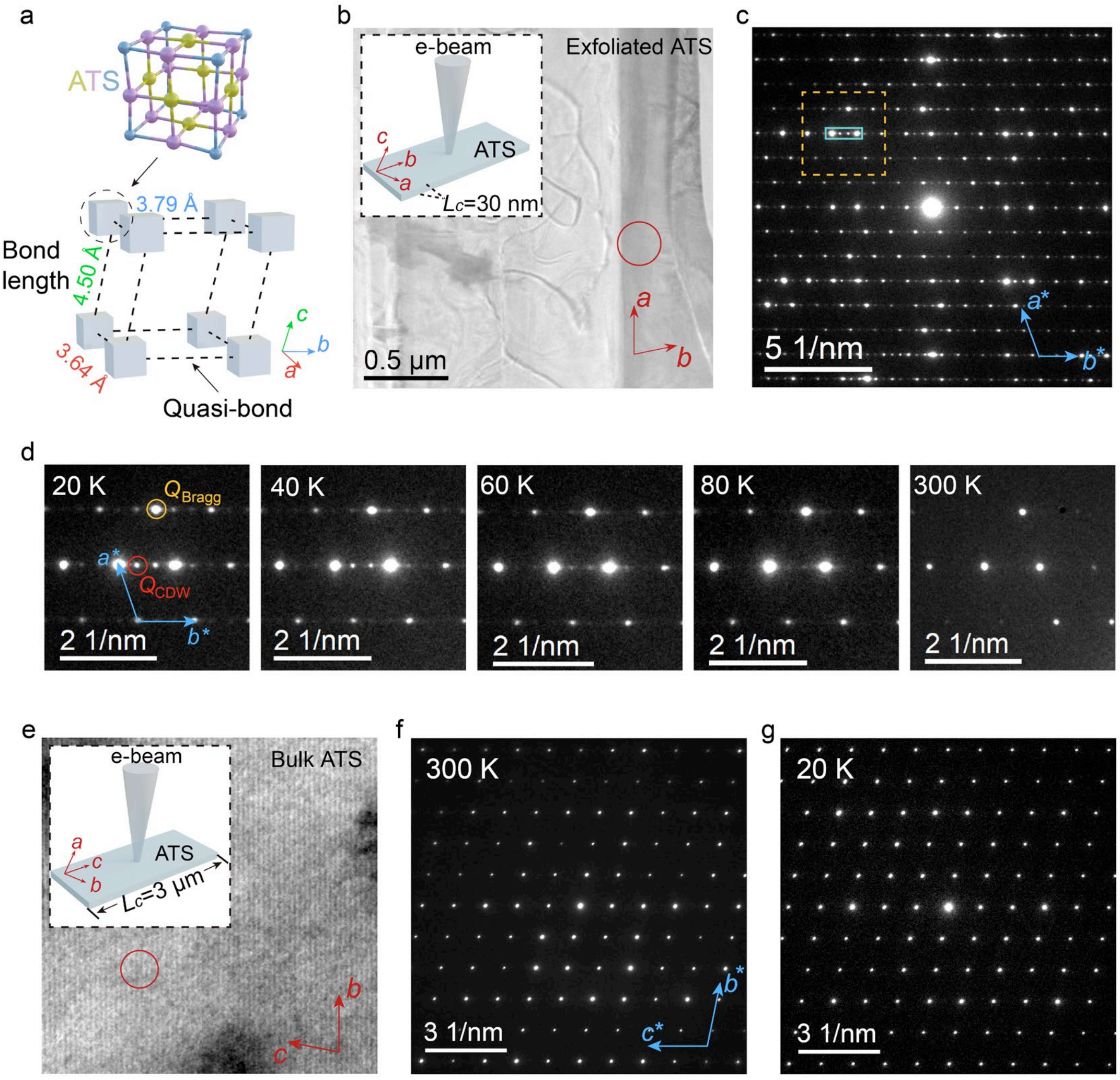

**Fig. 1 | Superatomic lattice reconstruction.** **a** Schematic of the $Au_6Te_{12}Se_8$ superatom connected through inter-cluster quasi-bonds. Bond lengths along three axes are labeled using the color same to the axes. Each unit cell contains one superatom. **b** Real-space image of an exfoliated thin flake. Selected-area electron diffraction (SAED) is performed in the red-circled region, with the electron-beam oriented perpendicular to the *ab*-plane. Inset illustrates the experimental geometry. **c** SAED pattern at 20 K revealing additional diffraction spots at 1/3 $b^*$. **d** Temperature evolution of CDW peaks acquired from the orange dashed square shown in (**c**). Superlattice spots broaden at 80 K and disappear by 300 K. **e** Focus ion beam (FIB) fabricated thick flake along the *c*-axis. SAED (red circle) was acquired with the beam parallel to the *bc*-plane (inset). **f**, **g** SAED patterns at 300 K (**f**) and 20 K (**g**), demonstrating no low-temperature superlattice formation.

The hallmark signature of CDW emerges in diffraction experiments, as CDW transitions simultaneously break symmetry in both electronic and lattice systems[34]. Using He-TEM, we uncover an interesting thickness dependence of CDW in $Au_6Te_{12}Se_8$. We first examine exfoliated flakes (~30 nm) with the electron beam oriented normal to the *ab*-plane (Fig. 1b inset). The red circle in Fig.1b indicates the area for selected area electron diffraction (SAED) on the exfoliated flake. Figure 1c presents the SAED pattern at 20 K, where the stronger diffraction spots correspond to the Bragg lattice. Notably, two weaker periodic diffraction spots appear between every two main spots. The intensity profile (shown in Supplementary Fig. 1) reveals that these spots are located at $1/3b^*$. This $a \times 3b$ reconstruction unambiguously confirms CDW formation. Temperature-dependent measurements in Fig. 1d (yellow boxed region in Fig. 1c) show the superlattice spots progressively broadening from 40 K to 80 K, completely vanishing by 300 K.

To probe thickness dependence, we fabricated a bulk-like flake (0.03 × 5 × 3 μm along $a \times b \times c$ axes) using FIB milling as shown in the inset of Fig. 1e. The beam is oriented normal to the *bc*-plane to ensure sufficient electron beam transmission. SAED pattern acquired at the red-circled region in Fig. 1e reveal a striking contrast. Unlike thin flakes, no superlattice spots emerge along either *b* or *c* directions at 20 K (Figs. 1f, g), despite clear resolution of fundamental Bragg peaks. This absence of low-temperature reconstruction in the bulk sample provides hints for the dimensionality-driven CDW.

**Visualization of transverse Peierls transition induced CDW**

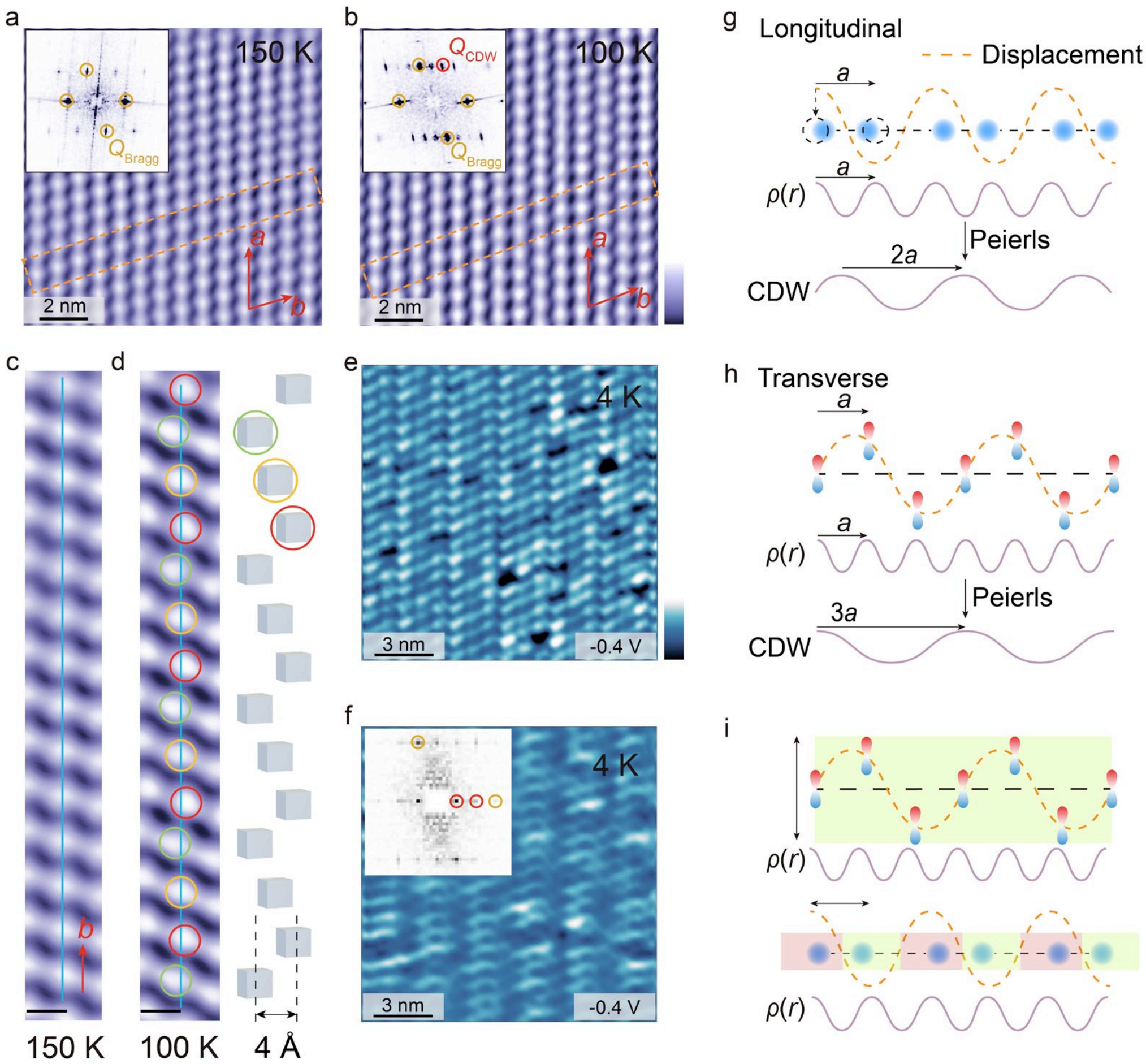

**Fig. 2 | Direct visualization of transverse Peierls transition induced CDW. a**, **b** STM topographies and corresponding fast Fourier transform of $Au_6Te_{12}Se_8$ at 150 K and 100 K. The set-points are -2.6 V, 100 pA and -2.2V, 100 pA, respectively. Bright spots represent individual superatoms rather than atoms due to delocalized valence electrons within superatoms. Bragg peaks of the main structure and the additional superlattice peaks are marked by yellow and red circles in the insets, respectively. The additional spots, appearing at lower temperatures, are located at (0, 1/3). Scale bars are 2 nm. **c**, **d** Enlarged STM topographies along the *b*-axis, as indicated by the orange dashed rectangles in (**a**) and (**b**). Blue lines along the *b*-axis serve as reference. Adjacent superatoms are color coded, showing clearly displacements along the *a*-axis. Schematic illustrations are displayed on the right of STM images with blue cubes representing superatoms. The CDW displacements reaches ~ 4 Å. Scale bars are 1 nm. **e** Topography of $Au_6Te_{12}Se_8$, showing the trimer-based transverse distortion pattern. **f**. d$I$/d$V$ map acquired simultaneously at a bias of -0.4 V with a set point of 100 pA. Inset is the FFT of the d$I$/d$V$, clearly revealing the 3$a$ periodicity in the electron system. **g** A sketch of the conventional CDW. The orange dashed line indicates the amplitude of the distortion, which is parallel to the CDW wavevector $\boldsymbol{Q}^*$. Purple lines indicate the charge density integrated in the green box area before and after the Peierls transition, which show 2$a$ period after transition. **h** A sketch of transverse Peierls transition with three-multiple modulation. Green boxes indicate the integral region of the charge density. $\rho(\mathbf{r})$ is display 3$a$ period after the Peierls transition. **i** Upper and lower panel show the case expanding the integral region to include whole lattice parameter for transverse and longitudinal Peirls transition, respectively. Both show no modulation, dashed yellow curve indicate the distortion.

Due to vibrational limitations in cryogenic He-TEM that preventing real-space imaging across the CDW transition, we employed STM to directly probe the lattice reconstruction. Figures 2a and 2b display STM images acquired at 150 K and 100 K with a tip bias of -2.6 V and- 2.2 V, respectively. While the superatoms are clearly visualized, the delocalized valence electrons within each superatom prevent resolution of constituent atoms. Fast Fourier transform results are shown in the inset of corresponding image. The key difference is the appearance of additional spots at (0, 1/3) at 100 K, signaling the formation of a CDW superlattice with a periodicity of $a \times 3b$, in perfect agreement with our He-TEM diffraction results.

Strikingly, we discover that the superatomic CDW in $Au_6Te_{12}Se_8$ can be directly resolved in real space without Fourier analysis, a capability overlooked in prior study[29]. By analyzing individual superatom columns along the *b*-axis (Figs. 2c, d), we observe a dramatic temperature-dependent transition: while superatoms remain uniformly aligned at 150 K, they develop significant *a*-axis displacements. We measure the d$I$/d$V$ map at 4 K. Figure 2 e and f show the topography and d$I$/d$V$ at the bias of -0.4 V and the set point of 100 pA. Alongside the distortion of lattice same to higher temperature shown in Fig. 2b, the real-space charge modulation is clearly revealed by the FFT of d$I$/d$V$ in the inset of Fig. 2f, confirming that the observed transition is not merely a periodic lattice distortion, but is accompanied by a periodic redistribution of charge density. Besides, Anti-phase relationship between occupied and unoccupied electronic states is found in the d$I$/d$V$ at +0.8 V and –0.8 V shown in Supplementary Fig. 2, which aligns well with the characteristic signatures observed in other established CDW materials. We also perform the d$I$/d$V$ map at a series of bias shown in Supplementary Fig. 3. The modulation phase shifts systematically with the bias while the topographies remain unchanged, ruling out static intra-cube charge redistribution.

Two prominent features of the observed charge order are worth noting. The first one is the observed CDW is result from a transverse Peierls transition[35]. The transverse Peierls transition was recently proposed to be a spontaneous symmetry broken mediated by the condensation of transverse phonon modes, rather than the conventional longitudinal modes. To clarify it, we present a toy model in Figs. 2g, h comparing a conventional Peierls transition and a transverse one along the $a$-axis. The charge density $\rho(\mathbf{r})$ of the conventional Peierls transition state shows a modulation along the $b$-axis with period of $2a$, namely a CDW. In the transverse Peierls transition state, the situation is more nuanced. Fig. 2h displays a trimerized transition modulate $\rho(\mathbf{r})$ with period of $3a$. One may argue that integrating over the complete crystal lattice parameter show no modulation as shown in the upper panel of Fig. 2i. However, this is natural because the Peierls distortion occurs along the $a$-axis, and such integration smears out the real-space modulation. This is analogous to the conventional longitudinal CDW (lower panel of Fig. 2i), if one integrates along the $b$-axis (over the red and green regions), the CDW modulation would also be averaged out. Namely, it is not appropriate to integrate along the direction of the Peierls transition. A rigorous definition and classification of CDWs are provided in the Supplementary Note. While the transverse Peierls transition has been proposed to induce a few novel properties, whether it can generate a CDW as conventional Peierls transition remains unknown. Our observations and model provide positive answer.

To conclusively verify the transverse distortion, we statistic ally analyzed 16 superatomic chains along the $a$-axis, as detailed in Supplementary Fig. 4. The inter-chain distance yield identical values, ruling out longitudinal distortion along the $b$-axis. This transverse displacement relative to the CDW wave vector $Q_{CDW}$ unambiguous evidence the emergence of a transverse Peierls transition. The transverse Peierls transition is recently proposed to possess unique properties different from conventional one, including plentiful gyro-acoustic phenomena and coupling with polarized optical field[35]. While its existence has been hinted by recently reported transverse acoustic phonons softening in $EuAl_4$[36-38], experimental visualization in atomic systems has remained elusive. Our work not only demonstrates clear lattice modulations at the superatomic scale but also establishes $Au_6Te_{12}Se_8$ as an ideal platform to directly explore the possible gyro-acoustic phenomena raised by the transverse Peierls transition.

The other feature of the CDW is the displacement taking place on the whole superatom, indicating the incorporated phonon modes are the collective vibration modes of superatoms. The distortion of superatoms induced by the CDW is unexpectedly large.

As shown in Fig. 2d, three successive adjacent superatoms are marked with green, red and yellow circles. A solid blue guide line is drawn through the centers of the yellow circles. We illustrate the distance measured in the schematic model in the right panel. Detailed topographic height profiles of individual superatoms with the reference to the guide line can be found in Supplementary Fig. 5. The magnified displacement of ~ 4 Å enables, for the first time, direct real-space visualization of CDW reconstruction, which is 1-2 orders of magnitude larger than conventional atomic CDW[39-41].

The observation of the CDW pattern on the bulk crystal surface via STM appears to contradict our previous assignment of a dimensionality-induced CDW order. We attribute this to a surface effect. As detailed in the Supplementary Fig. 6, the termination of crystal periodicity at the surface results in the loss of quasi-bond, leading to shift of the outmost few layers, which therefore behave like freestanding layers. Since STM is highly sensitive to the topmost layer, our measurements primarily probe these distinctive surface properties. However, for the nonlocal measurements such as electrical transport, the outmost few layers impose negligible effect as shown in the Supplementary Fig. 7.

## Role of electron-phonon-coupling

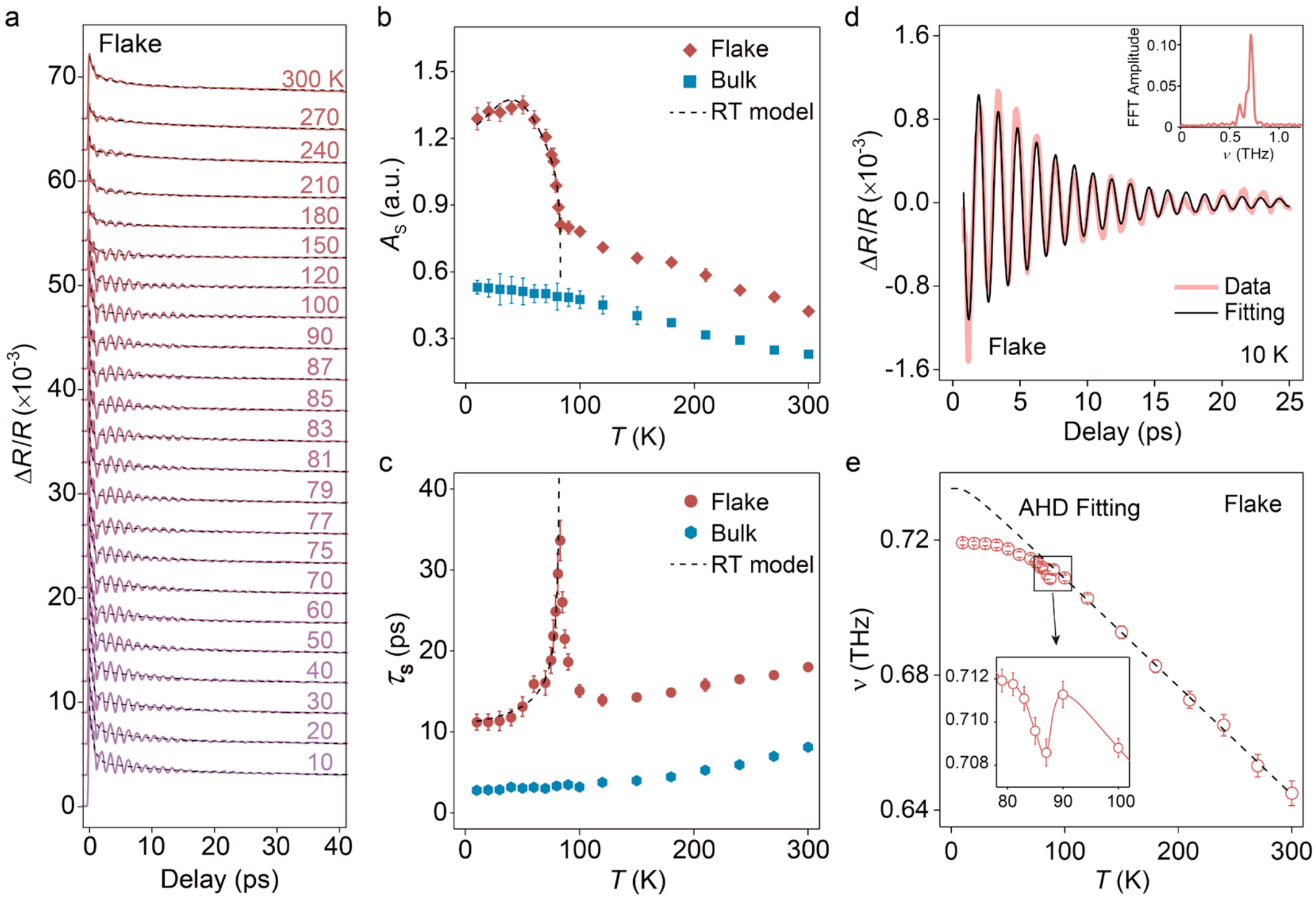

**Fig. | 3 Pump-probe measurement of the reflectance. a** Photoinduced reflectivity Δ*R*/*R* of a flake $Au_6Te_{12}Se_8$ sample with thickness of 48 nm measured from 300 K to 10 K. **b** and **c**, Amplitude ($A_s$) and relaxation time ($\tau_s$) of the decay process, extracted from the exponential fitting of the experimental Δ*R*/*R* data at different temperatures. Red and blue symbols represent the fitting parameters of the flake and bulk sample, respectively. Error bars represent the standard deviation of the fit. The black dashed curves are fits using the Rothwarf-Taylor model.

**d** Dashed line represents the fitting of the oscillations in $\Delta R/R$ of the flake sample at 10 K with the linear combination of two damped harmonic oscillators. Inset is the amplitude of the FFT of the oscillations. **e** Temperature-dependent frequencies of the fitted coherent phonon mode. The dashed guide line represents anharmonic decay (AHD) fitting, with deviations indicating the onset of CDW. Error bars represent the standard deviation of the fit.

The simultaneous electronic and lattice transitions observed in our system demand a thorough investigation of their interplay in CDW formation. CDW generally can be triggered by Fermi surface nesting[42], electron-phonon coupling (EPC)[43-46] or electronic interaction[47]. To investigate this, we perform ultrafast pump-probe measurements, which are highly sensitive to CDW-induced gap opening. The detailed setup is described in the Methods section and Supplementary Fig. 8. Figure 3a shows the time-dependent differential reflectivity $\Delta R/R$ of a flake sample with thickness (all thickness in next represent thickness along *c*-axis) of 48 nm, pumped with an 800 nm laser. $\Delta R/R$ can be fitted using two exponential decays and a constant term:

$$\Delta R/R = A_f e^{-t/\tau_f} + A_s e^{-t/\tau_s} + C.$$

Here, each exponential term represents a relaxation channel, and the thermal diffusion process is captured by the constant term $C$ with a cut-off delay time of 40 ps. This formula describes the relaxation process via quasiparticle scattering, where $A$ is proportional to the photoexcited quasiparticle density and $1/\tau$ represents the relaxation rate. The fitted curves for all temperatures are shown as dashed lines in Fig. 3a, and the fitted parameters $A_s$ and $\tau_s$ are plotted use red diamond and circles in Figs. 3b and c, respectively. As the temperature decreases to 87 K, $A_s$ abruptly increases and $\tau_s$ diverges, indicating a bottleneck effect caused by gap opening (the impact of $\tau_s$ divergence on the raw data see Supplementary Fig. 9). The relaxation process of the photoexcited quasiparticles after gap opening can be described by the Rothwarf-Taylor model[48,49], with the detailed fitting process provided in Supplementary Note and Supplementary Fig. 10. Using a BCS-like temperature-dependent gap function, we fit the quasiparticle density and obtain a CDW gap value of 14 meV. Interestingly, the measurements on bulk sample with identical set up show negative results of the gap opening (raw data of bulk see Supplementary Fig. 11), as shown by the blue squares in Fig. 3c. The vanish of CDW transition in bulk strongly indicates the previously observed CDW-like feature in transport measurements is actually originated from dimensionality.

We further analyze the oscillations[50] in the $\Delta R/R$ spectrum, which are the coherent phonon modes stimulated by the pump laser. Two same phonon modes can be found in the Raman spectrum in Supplementary Fig. 12. Fig. 3d shows the fitting of $\Delta R/R$ with linear combination of two damped harmonic oscillators[51], which can be also revealed by the fast Fourier transition shown in the inset. By choosing the rising edge of the raw data as the time zero point, the fitted initial phases of the two phonons are 1.47 (~π/2) and 4.65 (~3π/2), respectively. Given that we use sine function to fit the oscillations, both initial phases exactly lead to the cosine-like behavior, indicate the displacive excitation of coherent phonons (DECP) mechanism[52]. Here, we present the temperature dependence of the more prominent one in Fig. 3e, which depicts the phonon softening at the CDW transition temperature, as determined by fitting the oscillatory component to a damped harmonic oscillator. The other one phonon mode shows similar behavior, as shown in Supplementary Fig. 13. Their temperature dependence cannot be described by the conventional anharmonic decay (AHD) model[53] as indicated by the dashed line in Fig. 3e. The stimulated phonon modes collected here are close to $\Gamma$ point and away from $\boldsymbol{Q}_{\mathrm{CDW}}$, indicating a broad range of phonon softening differing from the Kohn anomaly. Combining with the absence of nesting of band structure at $\boldsymbol{Q}_{\mathrm{CDW}}$ according to the previous study[29], our result emphasizes that the EPC can play a crucial role in the transverse Peierls transition.

## Anomalous dimensionality effect of the CDW

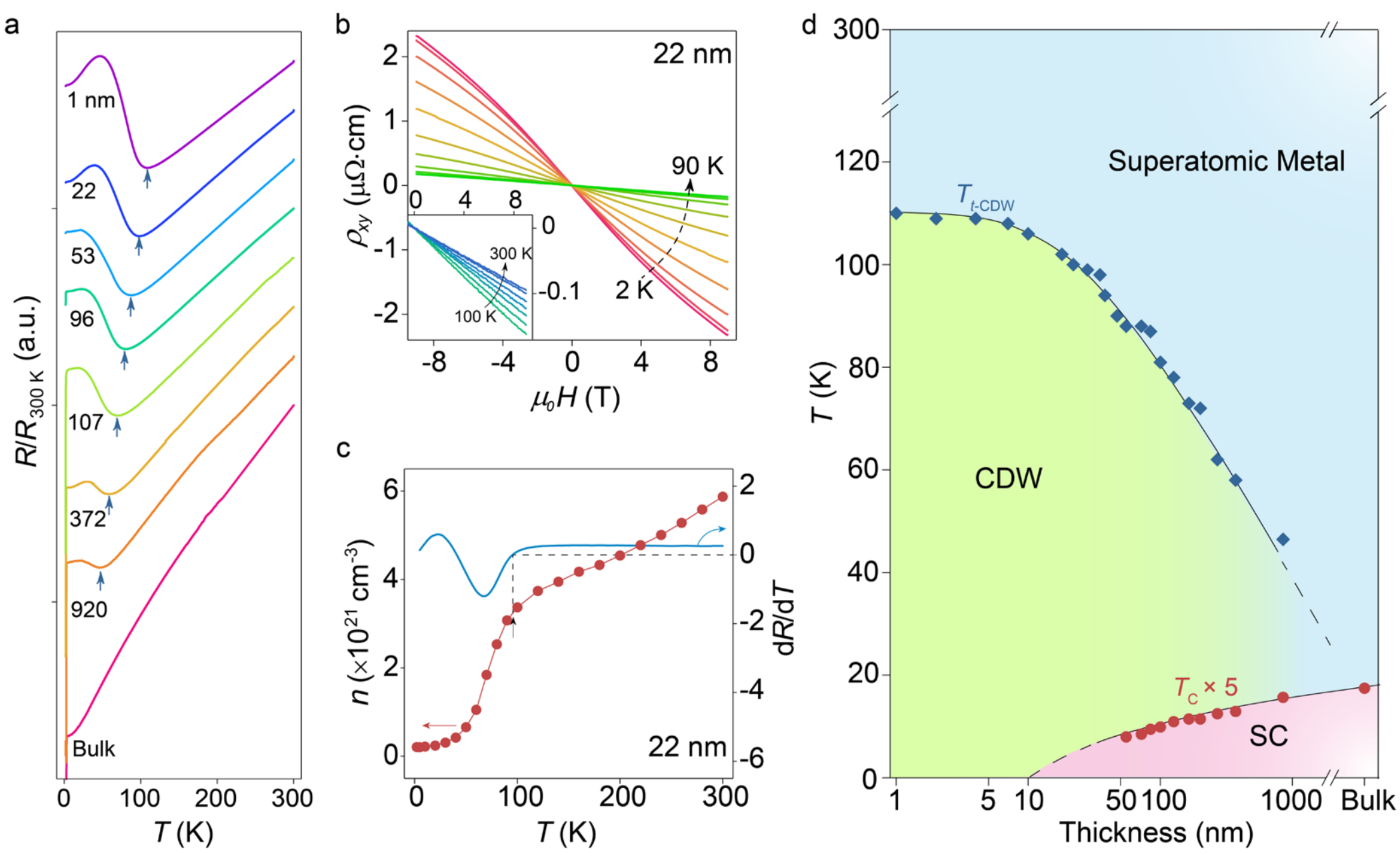

**Fig. | 4 Anomalous dimensionality effect of $Au_6Te_{12}Se_8$. a** Renormalized resistance of few exfoliated $Au_6Te_{12}Se_8$ sample with thickness varying from 2 nm to bulk. **b** $\rho_{xy}$ of the 22 nm sample at different temperatures from 2 to 90 K. Inset shows the data at 100-300 K. Carrier density can be extracted from the data and is shown in (c). **c** Derivative $\mathrm{d}R/\mathrm{d}T$ and the carrier density $n$ of the 22 nm sample. The resistance minimum (zero point of $\mathrm{d}R/\mathrm{d}T$) is consistent with the temperature where $n$ abruptly drops. **d** Phase diagram of the thickness dependency of the CDW and superconductivity in $Au_6Te_{12}Se_8$. The CDW emerges at samples with thickness of about 1 $\mu$m and its transition temperature continuously enhances with the dimensional reduction. A clear competition between the CDW and the superconductivity is shown.

Given that the CDW exclusively emerges in thin flakes, a systematic investigation of thickness dependence in $Au_6Te_{12}Se_8$ becomes crucial. Leveraging its van der Waals nature, we employed the $Al_2O_3$-assisstant exfoliation method to obtain flake samples with a wide thickness range. The normalized resistance $R/R_{300\,\mathrm{K}}$ of samples with typical thickness are shown in Fig. 4a. The data are shifted vertically for clarity. The bulk sample show no signature of any transition while all thin flakes depict a clear electronic instability by showing a resistance minimum. To clarify the resistance behavior, we perform the Hall measurements in a 22 nm flake sample as shown in Fig. 4b. The extracted carrier density $n$ is plotted in Fig. 4c together with the derivative of resistance $dR/dT$. The resistance minimum (zero point of $dR/dT$) is consistent with the temperature that $n$ drastically decreases which is a typical signature of gap opening, suggesting the resistance minimum can be specified as the CDW transition. Besides, the temperature of the resistance minimum of the 53 nm sample is consistent with the gap opening (87 K) of the 48 nm sample observed in pump-probe measurements, establishing the resistance minimum as a reliable CDW indicator.

The CDW transition temperature $T_{CDW}$, marked by arrows in Fig. 4a, reveals an unprecedented dimensional dependence. Strikingly, while undetectable in bulk samples, the CDW emerges at 45 K in micron-thick flakes and strengthens dramatically to 110 K as thickness approaches the "superatomic limit". This behavior stands in stark contrast to conventional 2D materials, where influence of dimensionality on CDWs typically become observable at ultrathin samples (<10 nm). The exceptional dimensional response originates from $Au_6Te_{12}Se_8$'s unique bonding architecture. Unlike traditional 2D materials with strong in-plane covalent bonds and weak interlayer van der Waals interactions, the superatomic lattice features weakly covalent quasi-bonds distributed uniformly along all three crystallographic directions. This three-dimensional bonding character results in the enhanced interlayer coupling, reduced electronic anisotropy, and ultimately, the observed giant dimensional enhancement of $T_{CDW}$, enabling the stabilization of CDW to much thicker region compared with conventional atomic 2D systems.

Additionally, while CDWs are typically suppressed by thermal fluctuation approaching the atomic limit, $Au_6Te_{12}Se_8$ exhibits the opposite trend. Only rare exceptions to this trend exist, typically arising from either enhanced electronic correlations or competing phases[54,55]. Given the CDW is relevant to the EPC rather than nesting in $Au_6Te_{12}Se_8$, we suggest that the existence of the competing superconducting phase should be crucial. Figure 4d summarizes transport data across all measured samples, revealing a clear competition between superconductivity and the CDW. The competing relationship between the superconductivity and the CDW is clearly revealed. As superconductivity and the CDW in $Au_6Te_{12}Se_8$ are both rely on EPC, which means the relationship between the phonon modes mediating superconductivity and superatomic distortions and their dimensionality effect may share a same origin, which we will discuss in detail later.

## Discussions

Superatoms generally form determined electronic structures, and consequently dictate the external bonding. In $Au_6Te_{12}Se_8$, the psudo-potential of our simple DFT calculations include total 186 electrons (Supplementary Fig. 14). The full filled bands below the Fermi level contains 184 electrons, forming closed shells in the molecular orbital (blue colored region). This leaves only 2 residual electrons per superatom, which originate from the *p*-orbitals of tellurium and constitute the superatomic valence electrons (red region). The inner electrons within the closed shells are highly stable and shared collectively by all the atoms inside the superatom. This explains why the STM imaging only resolves the superatomic structure rather than individual atoms. The 2 residual *p*-orbital electrons are responsible for forming the unique covalent-like quasi-bonds between adjacent superatoms. Furthermore, the CDW is mediated by the condensation of the transverse phonon modes, which leads to the transverse distortion of superatoms. The overlap integral of *p*-orbital electron is more sensitive to shear distortion than the breathing distortion due to the deconstructive interference of the opposite phase. Thus, we use the *p*-orbital in the CDW schematics in Fig. 2. Although the $Au_6Te_{12}Se_8$ superatom possesses two residual valence electrons, it does not precisely mimic an alkaline earth metal. The crucial distinction lies in the highly directional nature of these *p*-orbitals, which satisfies the theoretical prerequisite for the formation of a transverse Peierls transition induced CDW.

The transverse Peierls transition is firstly proposed to be mediated by the condensation of the transverse phonon modes arising from Fermi surface nesting, similar to the longitudinal Peierls transition which leads to the CDW. The EPC is soon proved to driven the conventional CDW when the nesting is weak. So far, it is still unknown whether EPC could mediate a transverse Peierls transition induced CDW due to its scarcity. Based on our observation, the CDW in $Au_6Te_{12}Se_8$ is most likely mediated by the EPC because of the broad phonon softening behavior across the transition temperature observed in pump-probe measurements. As illustrate in Supplementary Fig. 15, the typical nesting-driven CDW will show softening phonon modes only at very narrow momentum range around the nesting vector $\boldsymbol{Q}_{CDW}$. On the contrary, EPC-driven CDW soften phonon modes at relatively broader range, representing the hallmark of EPC mechanism. The oscillations in our pump-probe measurements captured the coherent phonon modes near $\Gamma$, which is far from $\boldsymbol{Q}_{CDW}$ of $Au_6Te_{12}Se_8$. Therefore, a reasonable deduction is the softening extend to a $\boldsymbol{Q}$ near $\Gamma$, which indicates the EPC case. Nevertheless, a phonon spectrum collected by the inelastic X-ray could be helpful in further study.

The exceptionally large CDW displacement is also noteworthy. We propose that it may originate from the synergistic effect between EPC and the superatomic nature of $Au_6Te_{12}Se_8$ clusters. Firstly, the amplitude of a CDW distortion is generally enhanced by stronger EPC. Second, and more distinctively, this distortion occurs at the superatomic level. The superatomic lattice undergoes distortion while the internal structure of each individual superatom remains nearly rigid. This indicates that the phonon modes relevant to the CDW involve the collective vibration of entire superatoms relative to one another, rather than the vibrations of their constituent atoms. An intuitively understanding of the superatomic phonon modes is a harmonic oscillator similar to the atomic one but with much larger mass (due to the superatoms) and weaker stiffness (due to the quasi-bonds). Then according to the amplitude and frequency of the harmonic oscillator,

$$\omega = \sqrt{k / m},$$

$$A = \sqrt{2E / k},$$

where $k$ is the stiffness and $E$ is the energy, we conclude the prominent feature of the superatomic phonon modes, extreme low frequency and large amplitude. Although it is a naive scenario, we believe it captured the main features. It also indicates the phonon modes directly relevant to the CDW in superatomic material possibly located at very low frequency that beyond the detection limit of conventional method like Raman spectrum.

Generally, thermal fluctuation in lower dimensional materials tends to suppress the CDW transition temperature for atomic materials. To the best of our knowledge, the enhancement of CDW approaching the 2D limit is rare with only two exceptions, $NbSe_2$[54] and $CsV_3Sb_5$[55,56] due to their unique enhanced electron-phonon coupling and electron-electron repulsion, respectively. The CDW in bulk $NbSe_2$ is mainly driven by weak EPC (nesting), but the optical spectrum study reveals an anomalous strengthen of the EPC according to the mean field analysis, which ultimately leads to the enhancement of CDW. The driven force of the CDW in $CsV_3Sb_5$ transit from EPC to the electron correlation marked by the decoupling of the CDW with the phonon hardening in Raman spectrum. Since the CDW is relevant to the EPC, the suppression of superconductivity can be considered as a direct competition with the gradually strengthened EPC.

Another unpresented feature is the length scale at which dimensionality effects manifest. We observe prominent changes in both the CDW and the superconductivity at a thickness of hundreds of nanometers, a scale far beyond the conventional definition of "thin flakes" for atomic materials and typically considered the bulk region. This remarkable observation can be understanded by considering the extremely low electronic anisotropy of $Au_6Te_{12}Se_8$. The Mermin-Wagner theorem implies that dimensional suppression of long-range order is most effective in highly isotropic systems[57]. An opposite example is the monolayer $Bi_2Sr_2CaCu_2O_{8+\delta}$ showing same $T_c$ to the bulk[58], which is comprehensive because $Bi_2Sr_2CaCu_2O_{8+\delta}$ shows the highest anisotropy among all cuprates. In stark contrast, the superatoms in $Au_6Te_{12}Se_8$ are bonded by the quasi-bonds with resistivity anisotropy on the order of 1, apparently much lower than that of $NbSe_2$ (~240)[32] and $Bi_2Sr_2CaCu_2O_{8+\delta}$ (~$10^5$)[33], revealing a unique facet of superatomic materials.

To conclude, we report a transverse Peierls transition induced CDW in superatomic $Au_6Te_{12}Se_8$ that challenges conventional classification. It represents a distinct dimensionality-driven phase, persisting in micrometer-thick flakes while absent in bulk crystals. The superatomic architecture enables direct observation of colossal real-space displacements. Its competition with superconductivity implies coupling through dimensionally-tuned phonon modes, revealing a new mechanism for quantum phase competition. All these anomalous properties are related to its superatomic nature. This work establishes superatomic materials as a distinct platform for engineering mesoscale quantum phases.

## Methods

### crystals Synthesis

Single crystals of $Au_6Te_{12}Se_8$ are grown using the self-flux method. Starting materials with high purity Au powder (99.99%, Sigma Aldrich), Te powder (99.999%, Sigma Aldrich), and Se powder (99.99%, Sigma Aldrich) are stoichiometrically weighted and sealed in an evacuated silica tube in high vacuum and subsequently mounted into a muffle furnace. The furnace was heated up to 800 °C in 40 h and dwelled 10 h. Afterward, the furnace was slowly cooled down to 450 °C in 4 days and then shut down. Crystals collected from ingot are generally thin with lateral dimensions of 4 × 2 $mm^2$.

### STM measurements

Single crystals are cleaved in ultrahigh vacuum at room temperature and subsequently cooled for STM measurements, which are performed in a commercial variable-temperature STM (PanScan Freedom, RHK) operated in ultrahigh vacuum. Electrochemically etched polycrystalline tungsten calibrated on clean Au(111) surfaces is used for all our STM measurements tips. The STM topography is acquired in the constant-current mode, and the d$I$/d$V$ spectra are collected using a standard lock-in technique with a modulation frequency of 999.1 Hz. STM measurements are performed at variable temperatures for the phase transition characterization.

### STEM measurements

The SAED are performed on JEOL-2100F Schottky field-emission microscope equipped with a Gatan Quantum 965 imaging filter. A liquid Helium sample holder is used to cool the sample to the CDW temperature.

### Pump-probe measurement

An achromatic pump-probe system based on a mode-locked Yb:KGW laser system was employed. The laser pulses with wavelength of 800 nm and repetition rate of 50 kHz are generated by the optical parametric amplifier, which is divided into two beams. One served as the probe beam, and another passed through a BBO crystal to generate a 400 nm pump pulses. The pump and probe beams are focused onto the sample surface through a 20 × objective lens. The focused spot diameters of the pump and the probe pulse were 9 and 4 μm, respectively. The pulse duration, after passing through the cryostat window, is measured to be 50 fs. In the temperature dependence measurements, the pump and probe fluences on the sample were kept at 15 and 10 $\mu$J/$cm^{-2}$, respectively. The pump beam was modulated by a chopper with a frequency of 433 Hz and the reflected pump beam was filtered out. The reflected probe beams traversed through the same objective lens, received by a photo-diode detector and sampled by a lock-in amplifier to enhance the signal-to-noise ratio. The relative change of reflectivity $\Delta R(t)/R_0=[R(t)-R_0]/R_0$, where $r$ and $r_0$ are the reflectivity of the probe with and without the presence of pump pulses, respectively, was recorded as a function of the time delay between the pump and probe pulses.

**Fabrication and characterization of few-layer $Au_6Te_{12}Se_8$ devices**

We obtain few-layer $Au_6Te_{12}Se_8$ samples using an $Al_2O_3$-assisted exfoliation technique as described in ref. 55. Firstly, the bulk samples are exfoliated to expose a fresh surface using scotch tapes. Then, 60nm $Al_2O_3$ is deposited on the surface. Next, the stack was picked up by a thermal-released tape. Through this process, few layer samples are exfoliated benefit from the tight adhesion between sample surface and $Al_2O_3$. Subsequently, the stamp of $Al_2O_3/Au_6Te_{12}Se_8$ is released onto a piece of PDMS (polydimethylsiloxane) with the $Au_6Te_{12}Se_8$ side in contact with the PDMS surface. Sample thickness is determined using the atomic force microscope and the optical contrast. After that, selected samples are finally stamped assembly onto a silicon substrate coated with 300 nm $SiO_2$. The electrical transport devices are fabricated through a standard electron beam lithography (EBL) process and a metal deposition with 5/60 nm Cr/Au as the electrodes. Electrical transport measurements are performed in a Quantum Design physical properties measurements system (PPMS) equipped with Stanford Research Lock-in amplifier (SR 830).

**ACKNOWLEDGMENT**
We are grateful for the fruitful discussion with Prof. Xi Dai. This work was financially supported by the National Key Research and Development Program of China and National Natural Science Foundation of China (Grants No. 2021YFA1401800, 52250308, 52522201, 52525205, 2024YFA1611303 and 52302010). This work was supported by the Synergetic Extreme Condition User Facility (SECUF, https://cstr.cn/31123.02.SECUF). A portion of numerical computations were carried out at the Hefei Advanced Computing Center.

## Author Contributions

T.Y., J.G., and X.C. conceived the idea and provided project supervision. B.S. was responsible for the majority of sample syntheses, device fabrications, measurements, and analysis. B.S. performed the transport measurements. S.S. conducted the LHe TEM measurements. X.B., L.W., and B.S. carried out pump-probe spectroscopy. S.X., Z.C., and X.C. performed the STM measurements. J.D. undertook the theoretical calculations. B.S. and T.Y. analyzed the data and wrote the paper, with contributions from all authors.